\documentstyle[12pt,epsf]{article}

\newcommand{\beq}{\begin{equation}}
\newcommand{\eeq}{\end{equation}}
\newcommand{\bea}{\begin{eqnarray}}
\newcommand{\eea}{\end{eqnarray}}
\newcommand{\rar}{\rightarrow}
\newcommand{\lan}{\langle}
\newcommand{\ran}{\rangle}
\newcommand{\cF}{{\cal F}}
\newcommand{\cG}{{\cal G}}

\begin{document}

\begin{flushright}
\vbox{\begin{tabular}{l} MADPH-96-974  \\
	FERMILAB-PUB-96/418-T \\
	February 1997
	\end{tabular}}
\end{flushright}
\begin{center}
\vspace{+1cm}
\Large
{\bf On the bounds for the curvature and higher derivatives
 of the Isgur-Wise
function} \\
\vskip 0.5cm
\large
M.G. Olsson      \\
{\small \em Department of Physics, University of Wisconsin, Madison,
	\rm WI 53706} \\
\vspace*{+0.2cm}
	Sini\v{s}a Veseli \\
\vskip 0.1cm
{\small \em Fermi National Accelerator Laboratory, P.O. Box 500, Batavia,
	\rm IL 60510}
\end{center}
\thispagestyle{empty}
\vskip 0.7cm

\begin{abstract}
We discuss constraints imposed on the zero-recoil curvature and higher
derivatives of the Isgur-Wise function by a general quark model. These
constraints are expressed as bounds for a given slope parameter, and compared
with those based upon analyticity properties of QCD spectral functions. Our
results also indicate that in the analysis of the experimental data for
semileptonic $B\rar D^{(*)}$ decays it may be important to include at least
the third term in the form factor expansion about the zero recoil point.
\end{abstract}

\newpage
\section{Introduction}

It has been widely recognized for some time that heavy quark symmetry
(HQS) \cite{isgur,reviews} enormously simplifies the analysis
of the semileptonic $B\rar D^{(*)}$ decays. The six form factors
needed for the description of these decays are in the heavy quark limit
reduced to a single unknown form factor, the Isgur-Wise (IW) function
$\xi(\omega)$, where $\omega = v_{B}\cdot v_{D^{(*)}}$ is
the product of the four-velocities of the two mesons. Furthermore,
HQS also provides us with a prediction for the normalization
of the universal form factor at the zero-recoil point, i.e. $\xi(1)=1$.
As a consequence of that, normalizations of the physical form factors
$\cG (\omega)$ (for $B\rar D$ decay) and $\cF (\omega)$
(for $B\rar D^{*}$ decay),\footnote{In the absence of symmetry breaking
corrections these form factors would coincide with the IW function
$\xi(\omega)$.} are determined up to
radiative and power corrections. Therefore, by extrapolating
the experimental data for the differential decay rate to $\omega = 1$
one can obtain an accurate measurement of the Cabibbo-Kobayashi-Maskawa
parameter $|V_{cb}|$. The decay $B\rar D^{*} l\bar{\nu}_l$ is
ideally suited for this purpose \cite{neubert}. It is experimentally
clean mode, and  the decay rate at zero recoil is
is protected by Luke's theorem against first order $1/m_{Q}$
corrections \cite{luke}.

This analysis has already been performed by several experimental
groups \cite{albrecht}-\cite{buskulic}. In \cite{albrecht} and
\cite{abreu} the fit to the data assumed a linear form for
$\cF(\omega)$, i.e. $\cF(\omega)=\cF(1)[1-
\hat{a}(\omega-1)]$,\footnote{In
\cite{albrecht} several other parametrizations for $\cF(\omega)$
were also used, but all with only one degree of freedom.}
while in \cite{barish} and \cite{buskulic} fits with
quadratic form of the hadronic form factor were also attempted,
but with the conclusion that with the existing data samples
it was not possible to distinguish between the linear parametrization
and those with more degrees of freedom. It should be obvious
that retaining only the first term in the expansion will
inevitably lead to an underestimate of the slope parameter.
This point was made some time ago by Burdman \cite{burdman}, who
included the curvature (quadratic) term in a two parameter analysis.
Because of the statistical uncertainty introduced in a two parameter
fit, it is clearly important to obtain some
theoretical insight about the expansion parameters, in order to guide the
extrapolation to $\omega = 1$.

This issue has already been addressed in several papers
\cite{rafael}-\cite{boyd2}, by employing analyticity properties of
QCD spectral functions and unitarity. The resulting bounds proved to
be weak due to the presence of the $\Upsilon $ poles
below the $B\bar{B}$ threshold (or possible $B_{c}$ states below the
$BD^{(*)}$ threshold in \cite{boyd2}).
In the most recent work \cite{caprini2} Caprini and Neubert (CN)
improved bounds for the zero-recoil slope and curvature (i.e. the
second term in the expansion) of $\cF(\omega)$
and $\cG(\omega)$ by identifying a specific $B\rar D$ form factor
which does not receive contributions from the ground state $B_{c}$
poles. These authors have derived constraints between the slope and curvature
of that form factor using analyticity properties
of QCD spectral functions and unitarity, and then used heavy quark symmetry
to relate these results to corresponding constraints for $\cF(\omega)$
and $\cG(\omega)$.

In this letter we discuss an alternative approach for obtaining
allowed regions for the curvature and higher derivatives of
the IW function $\xi(\omega)$ for a given value of the slope.
Our results are obtained in the heavy quark limit and in the valence quark
approximation, with some physical input about the shape of the heavy-light
meson wave function. An advantage of the method is that, given
the above assumptions, constraints can be obtained not
only for the curvature, but also for any higher terms in the expansion.
Even though we do not take symmetry breaking corrections into
account, we believe that our results may also shed some light
for guiding the experimental extrapolation to $\omega=1$, especially
for estimating the higher order terms in the form factor expansion.

The rest of the paper is organized as follows: in Section
\ref{iwff} we present the general valence quark model expressions
for the IW function,
and also for the particular terms in its expansion about
$\omega = 1$. In Section \ref{bounds} we show how to extract
bounds on the higher expansion parameters, if the slope of the
IW function is given. Results are discussed in Section \ref{results}, and
conclusions are presented in Section \ref{conc}. In particular, we
conclude that in the case of $B\rar D^{(*)}$ semileptonic decays
an expansion of the IW function about the zero recoil point
is converging slowly for $\omega$ close to the maximum
velocity transfer $\omega_{max}$.\footnote{For $B\rar D^{*}l\bar{\nu}_{l}$
$\omega_{max}\simeq 1.5$, and for $B\rar Dl\bar{\nu}_{l}$ decays
$\omega_{max}\simeq 1.6$.}

\section{IW form factor in a general quark model}
\label{iwff}

In the valence quark approximation
the expression for the IW function describing the  $B\rar D^{(*)}$
transitions, is  given in terms of the $S$-wave  radial wave function $R(r)$
and energy  $E$ of the light degrees of freedom in the ground state
heavy-light meson
\cite{sadzikowski}-\cite{veseli},
\beq
\xi(\omega) = \frac{2}{\omega+1} \lan j_{0}(k r) \ran\ ,
\label{xi}
\eeq
where $j_{0}$ is the spherical Bessel function and
\beq
k = 2 E \sqrt{\frac{\omega-1}{\omega+1}}\ .
\eeq
The expectation value $\lan F(r) \ran $ is given by
\beq
\lan F(r) \ran = \int_{0}^{\infty} r^2 dr |R(r)|^{2} F(r)\ .
\eeq
We define expansion of $\xi(\omega)$ around $\omega = 1$ as
\beq
\xi(\omega) = 1-a(\omega-1)+b(\omega-1)^{2}-
c(\omega-1)^{3}+d(\omega-1)^{4}+\ldots\ .
\label{exp}
\eeq
Using (\ref{xi}) it is straightforward
to find expressions for the slope $a$ and higher order terms in
(\ref{exp}) \cite{olsson,veseli}.
We list here the first four
terms:
\bea
a &=& \frac{1}{2}+\frac{1}{3}E^{2}\lan r^{2}\ran\ ,
\label{a}\\
b &=& \frac{1}{4} + \frac{1}{3}E^{2}\lan r^{2}\ran
+\frac{1}{30}E^{4}\lan r^{4}\ran\ ,
\label{b}\\
c &=& \frac{1}{8} + \frac{1}{4}E^{2}\lan r^{2}\ran
+\frac{1}{20}E^{4}\lan r^{4}\ran +\frac{1}{630}E^{6}\lan r^{6}\ran\ ,
\label{c}\\
d &=& \frac{1}{16} + \frac{1}{6}E^{2}\lan r^{2}\ran
+\frac{1}{20}E^{4}\lan r^{4}\ran +\frac{1}{315}E^{6}\lan r^{6}\ran
+\frac{1}{22680}E^{8}\lan r^{8}\ran \ .
\label{d}
\eea
Note that all quantities are positive definite.
{}From (\ref{a}) one finds $E^{2}\lan r^{2}\ran = 3(a-\frac{1}{2})$,
which can be used to reexpress (\ref{b}), (\ref{c}) and (\ref{d})
in terms of $a$ as
\bea
b &=&   \frac{1}{4}+(a-\frac{1}{2}) +
\frac{3}{10}\beta(a-\frac{1}{2})^{2}\ ,
\label{b2}\\
c &=& \frac{1}{8}+\frac{3}{4}(a-\frac{1}{2}) +
\frac{9}{20}\beta(a-\frac{1}{2})^{2} +
\frac{3}{70}\gamma(a-\frac{1}{2})^{3}\ ,
\label{c2}\\
d &=& \frac{1}{16}+\frac{1}{2}(a-\frac{1}{2}) +
\frac{9}{20}\beta(a-\frac{1}{2})^{2} +
\frac{3}{35}\gamma(a-\frac{1}{2})^{3}+
\frac{1}{280}\delta(a-\frac{1}{2})^{4}\ .
\label{d2}
\eea
Here, we defined dimensionless quantities
\bea
\beta &=& \frac{\lan r^{4}\ran }{\lan r^{2}\ran^{2}}\ ,
\label{beta}\\
\gamma &=& \frac{\lan r^{6}\ran }{\lan r^{2}\ran^{3}}\ ,
\label{gamma}\\
\delta &=& \frac{\lan r^{8}\ran }{\lan r^{2}\ran^{4}}\ .
\label{delta}
\eea
{}From (\ref{a}) one can see that $a\geq 1/2$, and therefore
it is immediately evident that all of the above parameters
must be positive.\footnote{The bound $a\geq 1/2$
is a consequence of the prefactor $2/(\omega+1)$ in (\ref{xi}),
derivation of which is discussed in depth in \cite{veseli}, and
which is closely related to the valence quark approximation. The
relationship between the HQET sum rules and quark models
was investigated in \cite{veseli2}, where it was shown that
Bjorken \cite{bjorken,isgur2} and Voloshin \cite{voloshin}
sum rules can be used to construct a model which is self-consistent
in the heavy-quark limit.} However,
we can bound them more stringently in order to yield
more useful restrictions on the allowed ranges of $b$, $c$, etc.

\section{Bounds}
\label{bounds}

Since all terms which accompany $\beta$, $\gamma$ and $\delta$
in (\ref{b2})-(\ref{d2}) are positive definite, it should be obvious that
by restricting the allowed range for those parameters we also
restrict the range of allowed values of $b$, $c$ and $d$, for
a given slope parameter $a$. In other words, we want to
find $\beta_{min}(\beta_{max})$ so that
\beq
\beta_{min}\leq \beta \leq \beta_{max}\ ,
\eeq
and similarly for other parameters.

Without making any further assumptions about the particular
form of the heavy-light wave function, lower bounds for
$\beta$, $\gamma$ and $\delta$
can be obtained by considering the Schwartz-type inequality
\beq
\lan r^{2m}(r^{2n}-\lan r^{2}\ran^{n})^{2}  \ran \geq 0\ .
\eeq
For $m=0,1$ and  $n=1,2$ this yields
\bea
\beta_{min} &= & 1\ ,
\label{betamin} \\
\gamma_{min} &= & 1\ ,
\label{gammamin}\\
\delta_{min} &= & 1\ .
\label{deltamin}
\eea

In order to estimate the upper bounds we need some physical input.
Let us for the moment assume that the ground state wave function of the
light degrees of freedom in a heavy-light meson is given in the form
\beq
R(r)\propto \exp{(-r^{k})}\ ,
\label{wf}
\eeq
where $k > 0$. Note that any scale or normalization dependence in the wave
function is unimportant, since it would cancel out in the ratios
(\ref{beta})-(\ref{delta}). For example, $k=2$ would correspond
to the harmonic oscillator wave function, which is (with appropriate
scale parameter) often used as an approximation for the meson
wave function \cite{isgur3, scora}. Case $k=1$ corresponds to
the pure exponential, which seems to be favored by lattice
QCD \cite{duncan}.
Using (\ref{wf}) one can  find the expression
\beq
\frac{\lan r^{2 n}\ran }{\lan r^{2}\ran^{n}}
= \frac{\Gamma(\frac{2n+3}{k})[\Gamma(\frac{3}{k})]^{n-1}}{
[\Gamma(\frac{5}{k})]^{n}}\ ,
\eeq
which with $n=2,3$ and 4 yields $\beta$, $\gamma$ and $\delta$
for any $k$.
In Table \ref{tab}  we list the actual numbers for
several cases of interest. Clearly, as the value of $k$ gets smaller,
parameters $\beta$, $\gamma$, and $\delta$ get larger. Therefore,
the smallest acceptable value of $k$ will lead to the largest
acceptable values for our parameters.
Since lattice simulations \cite{duncan} support a pure exponential falloff
of the meson wave function $(k=1)$, one might argue that
choosing, for example, $k_{min}=1/2$ would leave more than enough room
for possible uncertainties in the specific
choice (\ref{wf}) of the long distance
behavior of the wave function. In that case we would have
(see Table \ref{tab})
\bea
\beta_{max} &\approx & 5.67\ ,
\label{betamax} \\
\gamma_{max} &\approx & 107.19\ ,
\label{gammamax}\\
\delta_{max} &\approx & 5091.38\ .
\label{deltamax}
\eea
We wish to emphasize here that
any value $k_{min}<1$ would  be acceptable as far as this part of the
analysis is concerned, because it is essentially guided only by the information
obtained from the lattice \cite{duncan}. By choosing $k_{min}$
closer to 1, one would obtain quite narrow range for
all parameters under consideration, as we shall see in the
following section.

\section{Results}
\label{results}

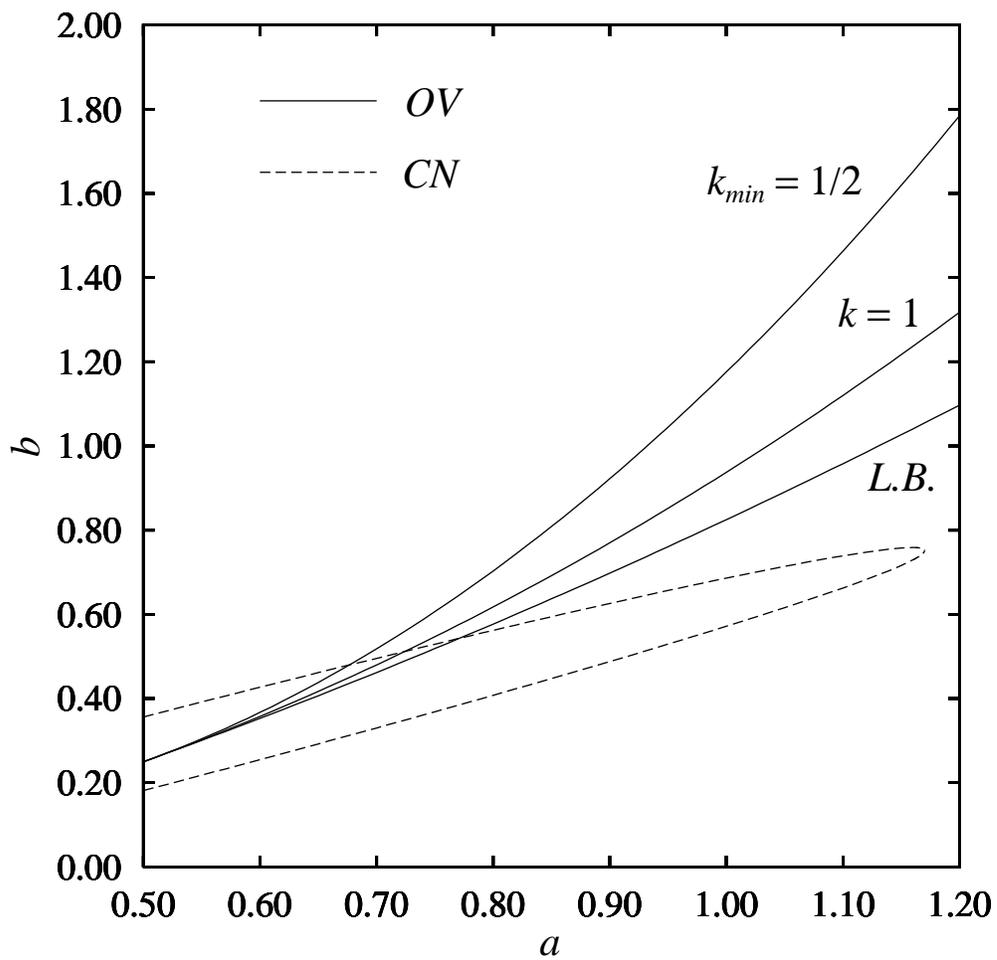
\begin{figure}
\caption{The valence quark model prediction for bounds on  the IW function
curvature $b$  in terms of its slope $a$ (full lines).
The line denoted by $k_{min}=1/2$ represents the upper,
while the line denoted by $L.B.$ represents the lower bounds.
The dashed line shows the result of the CN
analysis \protect\cite{caprini2}.}
\label{figb}
\end{figure}

Let us first discuss the second term  in the expansion
(\ref{exp}). In Figure \ref{figb} with full lines we show the
acceptable range for the curvature $b$ as a function of the slope
$a$, with a particular choice of $k_{min}=1/2$. The lower bound (denoted
by $L.B.$) follows from (\ref{betamin}), and the physically motivated
result with $k=1$ is shown as well. With dashed lines we further show
the result of the analysis performed in \cite{caprini2}, which
was also obtained in the heavy quark limit, but (unlike ours) includes
short-distance corrections. To be completely clear, we give here
(in our notation) the form factor definition used in \cite{caprini2}
\beq
\tilde{\xi}(\omega) = \tilde{\xi}(1)[
1-\tilde{a}(\omega-1) + \tilde{b}(w-1)^{2}+\ldots]\ .
\label{cnexp}
\eeq
To avoid confusion, we have used tilde
with the CN expansion parameters and form factor. When
the short-distance corrections ($\tilde{\xi}(1)\simeq 1.02$) are
neglected, (\ref{cnexp}) coincides with (\ref{exp}).\footnote{We
note here that perturbative corrections are expected to be at most 10-15\%.
For $\cF(\omega)$ in \cite{caprini2}
it was found that $\hat{a}\simeq\tilde{a}-0.06$ and
$\hat{b}\simeq \tilde{c}-0.06-0.06 \tilde{a}$, while for
$\cG(\omega)$ corresponding results were found to be
$\hat{a}\simeq\tilde{a}+0.02$ and
$\hat{b}\simeq \tilde{c}+0.01+0.02 \tilde{a}$.}

Since the CN ellipse shown in Figure \ref{figb} is rather narrow,
these authors also give the approximate relation between the slope
and the curvature as
\beq
\tilde{b}\simeq 0.72\tilde{a}-0.09\ .
\label{cnlin}
\eeq
On the other hand, our result with $k=1$, yields
\beq
b \simeq -0.06 + 0.25 a + 0.75 a^{2}\ ,
\label{bk1}
\eeq
which is within the CN bounds for values of $a$ smaller than about
0.7, and grows faster than (\ref{cnlin}) with increasing $a$.
{}From the Figure \ref{figb} one can see that given a value for the
IW function slope $a$, the valence quark model yields a
range for the curvature $b$ which is comparable in size to the
CN bounds, but with somewhat higher values for $b$ when $a$ is greater than
about 0.7. We remind the reader that (\ref{a}) requires $a\geq 0.5$.

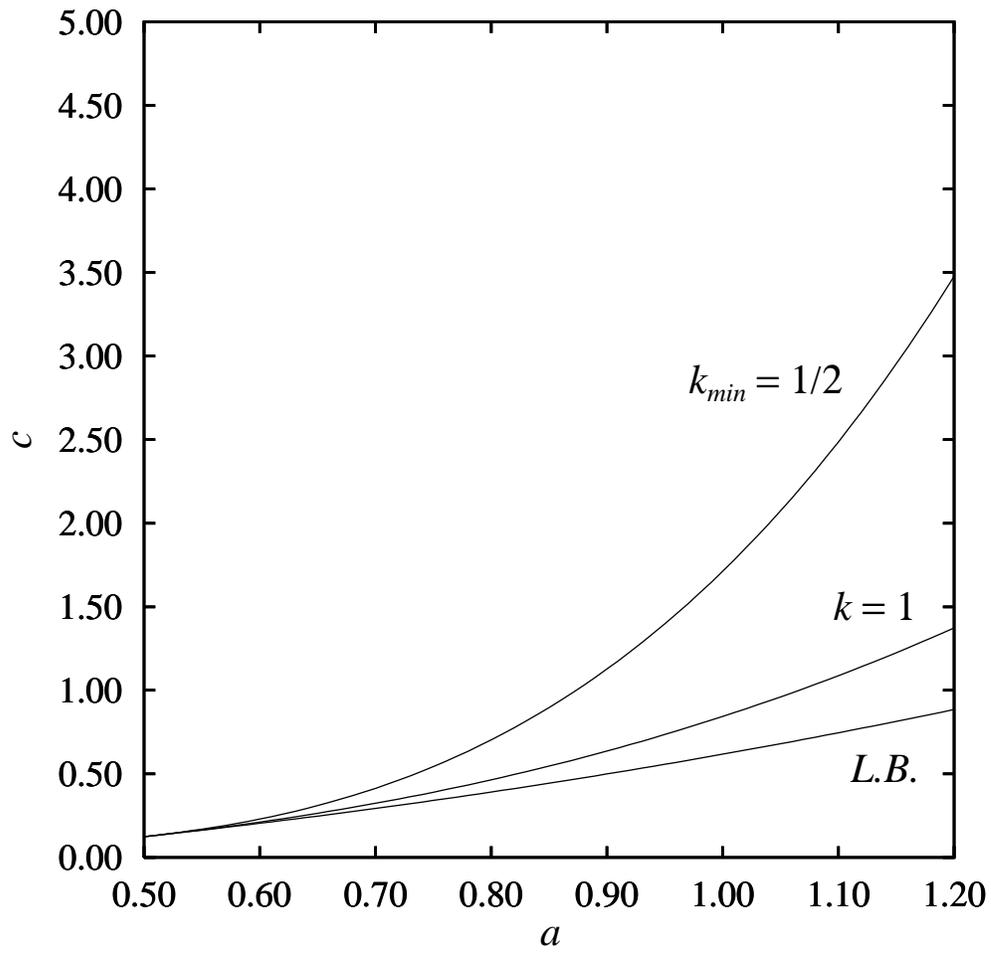
\begin{figure}
\caption{The valence quark model predictions for bounds
on the third term $c$ in the expansion of the
IW function (\protect\ref{exp}), in terms of the slope parameter $a$.
The line denoted by $k_{min}=1/2$ represents the upper,
while the line denoted by $L.B.$ represents the lower bounds.}
\label{figc}
\end{figure}

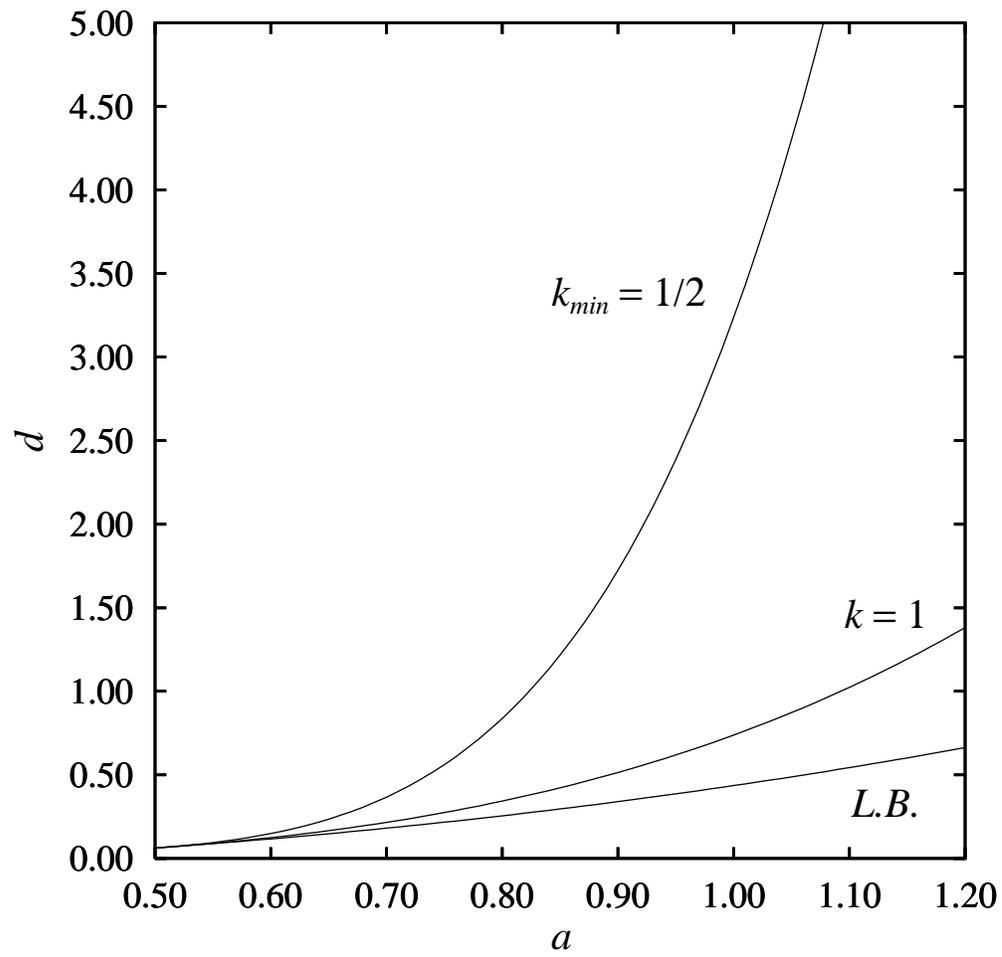
\begin{figure}
\caption{The valence quark model predictions for bounds
on the fourth term $d$ in the expansion of the
IW function (\protect\ref{exp}), in terms of the slope parameter $a$.
The line denoted by $k_{min}=1/2$ represents the upper,
while the line denoted by $L.B.$ represents the lower bounds.}
\label{figd}
\end{figure}

The CN approach is expected to break down for  higher
than the second terms in the expansion of the IW form factor. The
reason is the possible presence of the sub-threshold singularities which
are due to scalar $B_{c}$ resonances, or due to states of the form
($B_{c}^{(*)}+h$), where $h$ is a light hadron. We can, however, estimate
the acceptable range for those terms in the same way as we did for the
second term $b$. Results for the third and the fourth term ($c$ and $d$,
respectively), are shown in Figures \ref{figc} and \ref{figd}.
Naturally, for higher expansion parameters the uncertainty is increasing.
Nevertheless, if $k_{min}$ were close to 1, the range of acceptable values
for $c$ and $d$ would be quite narrow.
We note results for $c$ and $d$ obtained for the
physically motivated case of $k=1$, where we find
\bea
c &\simeq& -0.03  + 0.38 a^{2}  + 0.50 a^{3}\ ,
\label{ck1} \\
d &\simeq&
-0.01 - 0.03 a + 0.09 a^{2}  + 0.38 a^{3}  + 0.31 a^{4}\ .
\label{dk1}
\eea
One should also observe that
all expansion parameters are roughly of the same order of magnitude,
so that the only suppression for $n$-th order term in the expansion (\ref{exp})
is due to a factor of $(\omega-1)^{n}$. This fact is best illustrated  by
taking $a\simeq 1$ in (\ref{bk1}), (\ref{ck1}) and (\ref{dk1}), which leads to
$b\simeq 0.94$, $c\simeq 0.85$ and $d\simeq 0.74$.
Taking these values near the maximum velocity transfer in
$B\rar D^{*}l\bar{\nu}_{l}$
decays ($\omega_{max}\simeq 1.5$), we find from (\ref{exp})
\beq
\xi(1.5)\simeq 1 - 0.50 + 0.235  - 0.106 + 0.046-\ldots\  .
\eeq
Although the curvature term is large ($\sim 0.235$), the subsequent
terms are not negligible, and it is obvious that series
converges slowly for $\omega\simeq \omega_{max}$.

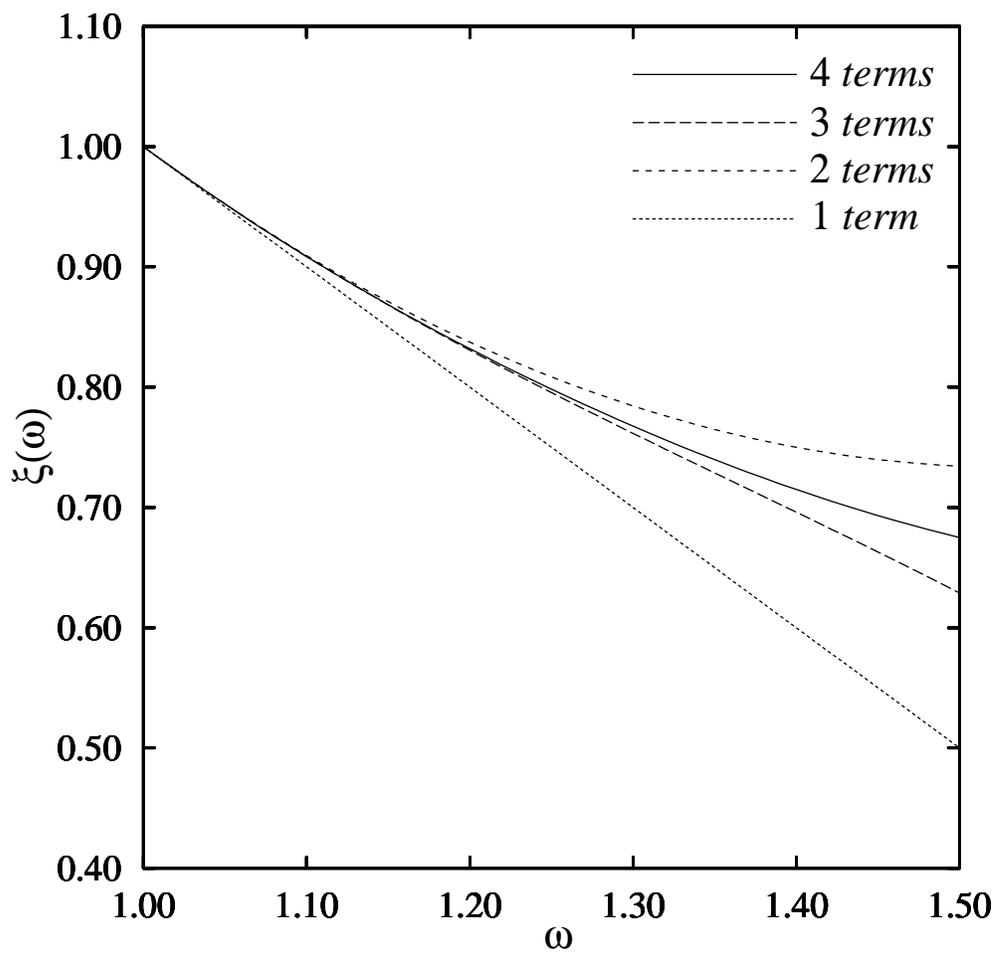
\begin{figure}
\caption{Illustration of the effects
of increasing the number of terms in the form factor
expansion about $\omega = 1$. We used $a=1$ and the physically
motivated case of $k=1$.}
\label{expfig}
\end{figure}

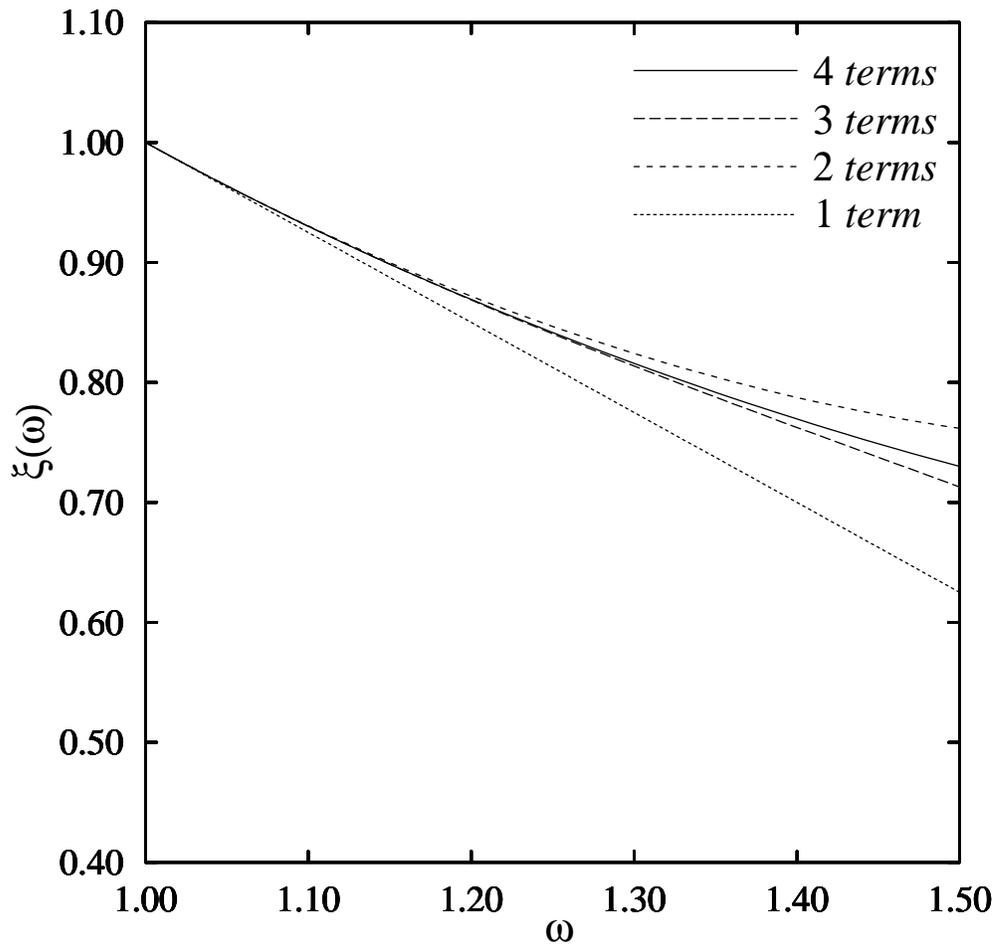
\begin{figure}
\caption{Same as in Figure \protect\ref{expfig} except with a
slope parameter  $a=0.75$. \hspace*{3cm}}
\label{expfig2}
\end{figure}

In order to show the effects of increasing of the number of terms in
the form factor expansion about $\omega = 1$, in Figure \ref{expfig} we show
what happens as we include one, two, three, and four terms in (\ref{exp}),
for $k=1$ and $a= 1$. Clearly, in this case keeping only the first two
terms ($a$ and $b$ non-zero) in (\ref{exp}) is an excellent
approximation to  $\xi(\omega)$
for $\omega\leq 1.2$. However, as $\omega$ increases,
the higher order terms  make a difference as far as the
shape of the form factor is concerned.  For  slope values
smaller than one the form factor convergence is better. This
is illustrated in Figure \ref{expfig2}, where we have used
a slope of $a=0.75$.
Nevertheless, in the analysis of the experimental data one has
to keep in mind that for larger values of the slope it may still
be important to include at least the third term in the
form factor expansion about $\omega = 1$.

\section{Conclusion}
\label{conc}

Within the framework of the general quark model, we have addressed the
issue of the bounds on curvature and higher
derivatives in the expansion of the Isgur-Wise function about
the zero recoil point. These terms are important  in experimental
extrapolation of the form factor towards $\omega = 1$. Except for
slopes in the range of $0.5$ to $0.7$, our
results indicate slightly larger curvature than the one obtained
by Caprini and Neubert \cite{caprini2}.  We also find that
including a third term in the form factor expansion about $\omega =1$
may be important in the analysis of the experimental data.

\begin{center}
ACKNOWLEDGMENTS
\end{center}
One of us (SV) would like to  thank
I. Dunietz for many fruitful discussions.
This work was supported in part by the U.S. Department of Energy
under Contracts No. DE-AC02-76CH03000 and DE-FG02-95ER40896.

\newpage
\begin{table}
\begin{center}
TABLES
\caption{Parameters $\beta$, $\gamma$ and $\delta$ for
$k=2$, $k=1$ and $k=1/2$.}
\label{tab}
\vskip 0.2cm
\begin{tabular}{|c|c|c|c|}
\hline
\hline
$k$ & $\beta$ & $\gamma$ & $\delta$ \\
\hline
$2$ & $\frac{5}{3} (\approx 1.67)$ & $\frac{35}{9}(\approx 3.89)$
	& $\frac{35}{3}(\approx 11.67)$ \\
$1$ & $\frac{5}{2} (=2.50) $ &$\frac{35}{3}(\approx 11.67)$
	& $\frac{175}{2}(= 87.50)$\\
$1/2$ & $\frac{715}{126} (\approx 5.67) $ &
$\frac{60775}{567}(\approx 107.19)$
	&$\frac{5773625}{1134}(\approx 5091.38)$ \\
\hline
\hline
\end{tabular}
\end{center}
\end{table}

\begin{figure}
\vspace*{+7cm}
\end{figure}

\clearpage
\newpage
\begin{figure}
\begin{center}
FIGURES
\end{center}
\end{figure}

\begin{figure}
\vspace*{+2cm}
\end{figure}

\clearpage

\begin{figure}[p]
\epsfxsize = 5.4in \epsfbox{fig1.ai}
\end{figure}

\begin{figure}[p]
\epsfxsize = 5.4in \epsfbox{fig2.ai}
\end{figure}

\begin{figure}[p]
\epsfxsize = 5.4in \epsfbox{fig3.ai}
\end{figure}

\begin{figure}[p]
\epsfxsize = 5.4in \epsfbox{fig4.ai}
\end{figure}

\begin{figure}[p]
\epsfxsize = 5.4in \epsfbox{fig5.ai}
\end{figure}

\end{document}